\documentclass[10pt]{article}
  \usepackage{PRIMEarxiv}
  \usepackage{listings}
  \usepackage{xcolor}
  \usepackage{amsmath}
  \usepackage{amssymb}
  \usepackage{caption}
  \usepackage[T1]{fontenc}
  \usepackage{tabularx,booktabs}
  \usepackage{ragged2e}
  \usepackage{multirow}
  \usepackage{subcaption}
  \usepackage{tikz}
  \usepackage{hyperref}
  \usetikzlibrary{positioning, arrows.meta}

  \hypersetup{
    colorlinks=true,
    linkcolor=black,
    citecolor=blue!50!black,
    urlcolor=blue!50!black,
  }

  \lstdefinelanguage{MLIR}{
    alsoletter={.,_},
    morekeywords=[1]{
      tensor.insert_slice, tensor.extract_slice,
      tensor.expand_shape, tensor.collapse_shape,
      tensor.empty, tensor.cast, tensor.pad,
      arith.constant, arith.addi, arith.muli,
      linalg.generic, linalg.matmul, linalg.fill,
      scf.for, scf.forall, scf.if, scf.yield,
      func.func, func.return, func.call,
      memref.alloc, memref.load, memref.store,
      affine.for, affine.if, affine.apply
    },
    morekeywords=[2]{into, from, outs, ins, attributes},
    morecomment=[l]{//},
    sensitive=true,
  }
  \lstdefinestyle{mlir}{
    language=MLIR,
    basicstyle=\ttfamily\footnotesize,
    keywordstyle=[1]\bfseries\color{blue!60!black},
    keywordstyle=[2]\itshape\color{black!60},
    commentstyle=\itshape\color{green!45!black},
    frame=single,
    framesep=4pt,
    rulecolor=\color{black!30},
    xleftmargin=1.5em,
    columns=fullflexible,
    keepspaces=true,
    showstringspaces=false,
  }

  \makeatletter
  \newcommand{\halfpointsize}{\@setfontsize{\onepointsize}{0.5pt}{0.5pt}}
  \makeatletter
  \newcommand{\onepointsize}{\@setfontsize{\onepointsize}{1pt}{1pt}}
  \makeatletter
  \newcommand{\twopointsize}{\@setfontsize{\twopointsize}{2pt}{2pt}}
  \makeatletter
  \newcommand{\threepointsize}{\@setfontsize{\threepointsize}{3pt}{3pt}}
  \makeatletter
  \newcommand{\lsrcsize}{\@setfontsize{\lsrcsize}{4pt}{4pt}}
  \makeatletter
  \newcommand{\fourpointsize}{\@setfontsize{\fourpointsize}{4pt}{4pt}}
  \makeatletter
  \newcommand{\fivepointsize}{\@setfontsize{\fivepointsize}{5pt}{5pt}}

  \AtBeginDocument{%
    }

\begin{document}
  \title{Reading AI Model Compilation in MLIR Through the Lens of Formal Theories}
  \author{Javed Absar \\
    Qualcomm Technologies International, Ltd. \\
    Cambridge, United Kingdom \\
    \texttt{javed.absar@gmail.com}}
  \date{}
  \maketitle

  \begin{abstract}
    Compiler infrastructures such as MLIR rest on a set of
    design principles: IR abstractions, interfaces,
    match-and-rewrite, flow analysis, type conversion, staged lowering,
    and so on.
    These concepts have proven themselves in practice.
    Good designs typically arrive through engineering knowledge, intuition and experience.
    Many of them, however, have correspondences in formal theory.
    MLIR's match-and-rewrite engine has correspondence to a \emph{term-rewriting
    system}~\cite{baadernipkow1998}; staged lowering has the structure of
    \emph{refinement calculus}~\cite{back1998}; and range analysis
    is grounded in \emph{abstract interpretation}~\cite{cousot1977,cousot1979}.
    Highlighting these correspondences is useful because each theory
    supplies vocabulary precise enough to discuss structural questions.
    Moreover, as coding agents lower the cost of implementation,
    good design and abstractions become the main concern~\cite{Lattner2026ClaudeCCompiler}.
    A coding agent can generate a pass, but it can only reason over the semantics
    the representation exposes. When essential structure is missing, the limitation is one
    of abstraction, not of implementation.
    The natural next question is how to design that substrate well.
    Well-chosen abstractions emerge from experience and intuition, but
    they often mirror concepts given a more precise treatment in formal
    theory. We argue that knowledge of these formal concepts clarifies
    what completeness means for a given abstraction, what the ideal
    design would be, and where practical trade-offs depart from it.
  \end{abstract}


  \section{Introduction}
  \label{sec:intro}

  In this paper, we touch on a number of formal theories and show
  correspondences with MLIR, and with compilers in general.
  Compilers for AI model compilation built using MLIR include
  Qualcomm's Hexagon-MLIR~\cite{hexagonmlir2026}, Intel's AI
  Compiler~\cite{golin2024}, IREE and TinyIREE~\cite{iree-github,tinyiree2022},
  PolyBlocks~\cite{polyblocks2026}, AMD's Aster~\cite{aster2026},
  NVIDIA's CUDA Tile IR~\cite{cudatile2026}, AXI4MLIR~\cite{axi4mlir2024},
  and the open-source Project Lighthouse~\cite{mlir_lighthouse}.
  MLIR~\cite{mlir2021,vasilache2022,lucke2024} simplifies their construction by
  providing infrastructure for defining new intermediate representations,
  dialects, transformations, pattern rewrites,
  verification, and extensible interfaces.

  Compilers are often organised as layered stacks of intermediate
  representations (IR)~\cite{clickir1995,halide2013,tvm2018}.
  A PyTorch program may pass through TorchDynamo
  and TorchInductor, be lowered to Triton, then to MLIR, and finally to
  LLVM IR before being emitted as machine code. Each layer has its own
  IR and its own mechanisms for lowering and optimisation. Each layer
  commits to more concrete detail, and the engineering decisions at
  every level are how to analyse and optimise the IR, and which
  abstractions to retain versus which to commit to concrete choices.

  These designs come either from formal theory or from experience
  with what works well. Good designs acquire a formal framework
  over time, and sometimes good formal theory is borrowed in
  implementation to make it sound. We do not wade into what came first.
  The important point is to highlight the correspondences.
  Figure~\ref{fig:correspondence} summarises the ones this paper
  discusses.

  \begin{figure}[htbp]
  \centering
  \renewcommand{\arraystretch}{1.4}
  \begin{tabular}{r@{\quad}c@{\quad}l}
  \toprule
  \textbf{Formal theory} & & \textbf{MLIR construct} \\
  \midrule
  term rewriting: $\ell \to r$
    & $\rightleftarrows$
    & \texttt{rewrite pattern}, \texttt{op-fold}, canonicalisation \\
  refinement calculus: $P_1 \sqsubseteq P_2$
    & $\rightleftarrows$
    & progressive lowering, type converter \\
  abstract interpretation: $\alpha,\,\gamma$
    & $\rightleftarrows$
    & \texttt{interval arithmetic}, \texttt{value bounds analysis} \\
  type theory: $\Gamma \vdash e : T$
    & $\rightleftarrows$
    & op / type / attr interfaces \\
  category theory: $f : A \to B,\; g \circ f$
    & $\rightleftarrows$
    & pass, pipeline (composition of passes) \\
  \bottomrule
  \end{tabular}
  \caption{Each formal framework on the left is read against an MLIR
    construct on the right that has at least some correspondence to it.
    The bidirectional arrows show the cross-pollination. Successful ideas
    developed in practice are sometimes followed by theory that
    generalizes and abstracts them, giving a more precise formulation.}
  \label{fig:correspondence}
  \end{figure}

  With the emergence of agentic code assistants in 2024 --- Claude Code
  (Anthropic)~\cite{santos2025claudecodeconfig} and so on
  --- the cost of \emph{writing} analyses, transformations, and passes is falling rapidly,
  and these agents have
  begun producing compiler-grade code~\cite{mateega2026idebench}. Chris
  Lattner's analysis of \emph{The Claude C
  Compiler}~\cite{Lattner2026ClaudeCCompiler} makes two observations
  relevant here: agent-generated compilers tend toward LLVM-shaped
  designs, because their training data is the history of compiler
  engineering; and the visible automation is of \emph{implementation},
  not of design. A coding agent can generate a pass, but
  it can only reason over the semantics the representation
  exposes. When essential structure is missing, the limitation is one
  of abstraction, not of implementation, and no amount of agent skill
  can recover what the IR has thrown away and does not represent.

  Taken together, these observations move the bottleneck of compiler
  engineering from writing passes to designing the substrate they
  run on. The natural next question is how to design that substrate well.
  Well-chosen abstractions emerge from experience and intuition, but
  they often mirror concepts given a more precise treatment in formal
  theory. We argue that these correspondences are useful as vocabulary
  for stating precisely what the code implements, what the ideal form
  of the abstraction would be, and what structural questions a designer
  should ask.

  As this paper is focused on abstraction, we order the
  formal theories by increasing distance from implementation:
  term-rewriting systems (TRSs) are easily related to a working
  compiler's pattern infrastructure, while category theory ---
  which appears later briefly --- is more abstract and gives an
  implementation-independent perspective on structure-preserving
  maps between dialects.
  We treat each theory as design vocabulary, not as hard foundation.
  Each formal framework illuminates structural questions a
  designer faces, and the value of the framework is in how crisply
  it lets that question be stated. 

  The paper covers a number of formal theories:
  \emph{term rewriting}
  (Section~\ref{sec:termrewriting}), the engine of pattern-based
  optimisation in MLIR; \emph{refinement calculus}
  (Section~\ref{sec:refinement}), which frames each lowering step as
  a contract-preserving derivation; \emph{abstract interpretation}
  (Section~\ref{sec:absint}), which characterises what information is
  preserved or lost at each abstraction boundary; and briefly,
  \emph{type theory}, \emph{category theory}, and
  \emph{operational/denotational semantics}
  (Section~\ref{sec:otherframeworks}), which offer complementary
  perspectives on well-formedness, structure-preserving maps, and
  program meaning.

  \section{Term Rewriting}
  \label{sec:termrewriting}

  Term rewriting~\cite{baadernipkow1998,meseguer2012} is a formal theory of
  directed equational reasoning. A term-rewriting system (TRS) is a
  set of rules $\ell \to r$ over a term algebra. A rule applies to a
  subterm matching $\ell$ (a \emph{redex}) by substituting $r$ for
  $\ell$ in context. Successive applications produce a reduction
  sequence $t_0 \to t_1 \to t_2 \to \cdots$. A term with no remaining
  redex is a \emph{normal form}. TRS theory asks two questions about
  such systems --- confluence and termination --- and uses machinery
  (Newman's lemma, critical pairs, the Knuth--Bendix procedure) for
  answering them. 

  \paragraph{Confluence.}
  A system is \emph{confluent} when all reduction orders converge
  to the same result. In other words, any two reductions of the same
  term reach a common descendant.

  \paragraph{Termination.}
  The strong form of the halting question is \emph{strong
  normalisation} (or termination): no infinite reduction sequence
  exists. The weaker form is \emph{weak normalisation}: from each
  term, \emph{some} reduction sequence reaches a normal form, even
  if other choices diverge. A confluent strongly-normalising system
  has a unique normal form for every term.

  \paragraph{Newman's lemma and critical pairs.}
  A result in term-rewriting is Newman's lemma~\cite{newman1942}:
  for a terminating system, \emph{local confluence} implies
  confluence. The value of this reduction is
  that proving confluence directly requires showing all possible
  divergences reconverge --- an infinite check in general. Newman's lemma
  says: it suffices to check that
  any two one-step rewrites of the same term can be joined by further
  rewrites (local confluence). Confluence guarantees that no matter
  which rule fires first, the system reaches the same final result ---
  the choice of rule application order doesn't matter. Rules that fire
  on disjoint subterms cannot interfere: each rewrite leaves the
  other's redex intact, so both can still fire and reach the same
  result. The only problematic case is when two rules compete for the
  same subterm. For a finite rule set there are only finitely many such
  overlaps, so checking local confluence becomes a finite enumeration
  --- checking that every \emph{critical pair} (a term where two rules
  overlap on the same subterm) is joinable. 
     
  \paragraph{Knuth--Bendix completion.}
  The Knuth--Bendix completion procedure~\cite{knuthbendix1970}
  uses critical pairs in the other direction. Given a terminating
  non-confluent system, it adds rules that close each open critical
  pair. If it succeeds, the result is a confluent terminating
  system whose normal forms decide the underlying equational
  theory. The procedure is mechanical: given an open critical pair
  whose two sides reduce to distinct normal forms $s'$ and $t'$, it
  orients the equality as a new rule $s' \to t'$ (or $t' \to s'$),
  choosing the direction so that the larger term under a fixed
  termination ordering rewrites to the smaller. This guarantees the
  enlarged system still terminates --- every new rule strictly reduces
  term size. The procedure then recomputes critical pairs for the
  enlarged rule set and repeats. It succeeds when all pairs close, and
  diverges when new rules keep generating new open pairs indefinitely.
  
  \begin{figure}[htbp]
  \centering
  \begin{subfigure}[b]{0.42\textwidth}
  \centering
  \begin{tikzpicture}[
    node distance=1.4cm,
    every node/.style={font=\small}
  ]
  \node (t) {$t$};
  \node[below left=of t] (u) {$u$};
  \node[below right=of t] (v) {$v$};
  \node[below=2.5cm of t] (w) {$w$};
  \draw[->, thick] (t) -- node[above left, font=\scriptsize] {$*$} (u);
  \draw[->, thick] (t) -- node[above right, font=\scriptsize] {$*$} (v);
  \draw[->, thick, dashed] (u) -- node[below left, font=\scriptsize] {$*$} (w);
  \draw[->, thick, dashed] (v) -- node[below right, font=\scriptsize] {$*$} (w);
  \end{tikzpicture}
  \caption{Confluence: any two reductions of $t$ can be extended to
    a common descendant $w$. Solid arrows are the given reductions,
    dashed the required ones; $*$ denotes zero-or-more steps.}
  \label{fig:confluence}
  \end{subfigure}
  \hfill
  \begin{subfigure}[b]{0.5\textwidth}
  \centering
  \vspace{1.6cm}
  \begin{tikzpicture}[
    node distance=1.8cm,
    every node/.style={font=\small}
  ]
  \node (a) {$a$};
  \node[right=of a] (b) {$b$};
  \node[right=of b] (c) {$c$};
  \draw[->, thick] (a) to[bend left=22] (b);
  \draw[->, thick] (b) to[bend left=22] (a);
  \draw[->, thick] (b) -- (c);
  \end{tikzpicture}
  \vspace{1.4cm}
  \caption{Non-termination and non-confluence under rules $a \to b$,
    $b \to a$, $b \to c$. The reduction $a \to b \to a \to b \to
    \cdots$ never halts; from $b$, one path reaches the normal form
    $c$ while another stays in $\{a,b\}$ forever, so the divergence
    has no common descendant.}
  \label{fig:nontermination}
  \end{subfigure}
  \caption{A confluent system (left), where any divergence can be
    reconverged, versus a system (right) where rules induce a cycle
    and a divergence with no common descendant. When a canonicalizer
    is built from rules of the second kind, the order of application
    decides whether it terminates and what normal form it produces.}
  \label{fig:trs-properties}
  \end{figure}
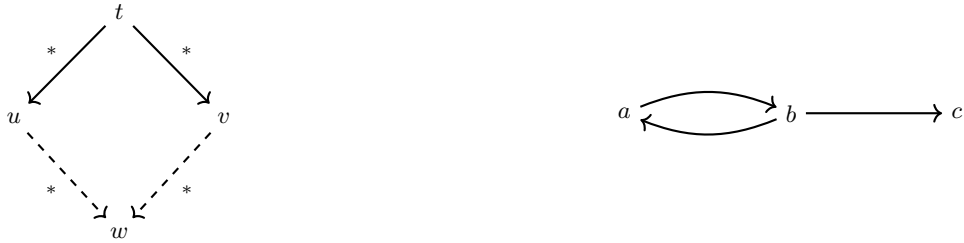

  \paragraph{Term Rewriting and MLIR.}
  Compilation of AI models in MLIR is largely a story of rewriting:
  transforming IR for optimisation, or lowering it from abstract
  dialects toward machine instructions (e.g.\
  \texttt{linalg} $\to$ \texttt{scf} $\to$ \texttt{cf} $\to$
  \texttt{llvm}). MLIR provides several rewriting mechanisms.

  The simplest is the \emph{IR walk}: a visitor-based API that
  allows the user to match and replace a subgraph in place.
  It is a manually-driven, one-pass rewrite
  with no fixpoint semantics --- the programmer controls traversal
  and rule selection explicitly.

  The \emph{greedy pattern rewrite driver}
  (\texttt{applyPatternsAndFoldGreedily}) is quite close to a TRS
  engine. A \texttt{RewritePattern} is a rule $\ell \to r$:
  its \texttt{matchAndRewrite} method defines the match (the LHS)
  and uses a \texttt{PatternRewriter} --- the mutation API --- to
  build the replacement (the RHS). The driver applies all registered
  patterns, folds and erases dead ops, and re-processes modified ops
  until no pattern fires (a fixed point) or a configurable iteration
  bound is reached. It has no rollback: once a pattern fires, its
  effect is committed. Neither confluence nor termination is
  guaranteed --- the order of pattern application may determine which
  fixed point is reached, and a poorly designed rule set can cycle
  until the iteration bound cuts it off.

  The \emph{dialect-conversion framework}
  (\texttt{applyFullConversion} / \texttt{applyPartialConversion})
  is structurally different. It uses \texttt{ConversionPattern} and
  \texttt{ConversionPatternRewriter}, applies patterns only to ops
  marked illegal, traverses top-down by dominance, and provides
  rollback: if a sequence of rewrites cannot complete, the IR is
  restored to its prior state. It also supplies a
  \texttt{TypeConverter} for coordinating type changes across the
  conversion boundary.
  From a TRS perspective, the dialect-conversion framework
  guarantees \emph{termination} (the source dialect is exhausted by
  construction --- legality rules forbid backward rewrites) and
  \emph{completeness} (every illegal op must have at least one
  conversion pattern, or the pass fails). Confluence, however, is
  not formally guaranteed: when multiple patterns can lower the same
  op, the framework picks one, and the dialect designer is
  responsible for ensuring that the alternatives produce equivalent
  results.
  Where the type conversion path has gaps, MLIR inserts an
  \texttt{unrealized\_conversion\_cast} as a placeholder, to be
  eliminated by a later conversion that can express the missing step.

  \paragraph{Confluence in practice.}
  Confluence is a property of the rule set, not of the framework.
  During dialect conversion (e.g.\ \texttt{arith} $\to$
  \texttt{llvm}) there is often a one-to-one correspondence between
  source and target ops, making the system trivially confluent ---
  if each op has exactly one rewrite, divergence is not possible.

  Within a dialect or across a set of cooperating dialects, the
  question is more interesting. Consider rewriting a
  \texttt{tensor<16x32xf32>} to
  \texttt{tensor<1x1x1x16x32xf32>}:

  \begin{lstlisting}[style=mlir]
%a = tensor.insert_slice %slice into %x [0, 0, 0, 0, 0]
                                        [1, 1, 1, 16, 32]
                                        [1, 1, 1, 1,  1]
       : tensor<16x32xf32> into tensor<1x1x1x16x32xf32>

%b = tensor.expand_shape %slice [[0, 1, 2, 3], [4]]
       : tensor<16x32xf32> into tensor<1x1x1x16x32xf32>
  \end{lstlisting}

  Both forms are useful in different contexts. In general, it is a
  dialect design question of what would lead to better code.

  In general, \emph{confluence} is a good design consideration.
  Unnecessary introduction of \emph{critical pairs} (where two rules
  overlap on the same subterm) only complicates
  the rewrite process and the writing of passes. A parallel is from
  LLVM where loops are often written out in normal form before
  loop-level optimisation passes.

  \paragraph{Canonicalization and normal forms.}
  Where TRS speaks of \emph{normal forms}, MLIR has
  \emph{canonicalization}: a per-op \texttt{fold} method plus a set
  of canonicalization patterns that simplify the IR toward a
  preferred shape. Canonicalization is not a formal normal form ---
  the rule sets are neither confluent nor strongly normalising in
  general --- but it serves the same engineering purpose: reducing
  the IR to a smaller form so that subsequent passes can pattern-match
  more easily.

  In addition to classical TRS-style rewriting, equality saturation
  using e-graphs has recently been explored in
  MLIR~\cite{merckx2026}. E-graphs represent many equivalent
  expressions simultaneously in equivalence classes and defer the
  choice of canonical form to a cost-driven extraction phase. Because
  all equivalent forms coexist, rule application order is irrelevant.
  The trade-off is space: equivalence classes can grow exponentially
  as rules fire, making the approach practical only for bounded
  subproblems or with aggressive saturation limits.

  \section{Refinement Calculus}
  \label{sec:refinement}
  While the previous section focused on the design of rewrite rules
  (confluence, termination), refinement calculus addresses the broader
  question: is the staged lowering process as a whole correct?

  Refinement Calculus~\cite{back1998,morgan1990} adds rigour to the
  question: how does one go from ``what the program should do'' to
  ``code that does it'' in small, individually verifiable steps?
  Specifications and implementations live in the same space --- both
  are programs --- and that space is ordered by a relation
  $\sqsubseteq$. One starts with a permissive specification at the
  top of the order and walks downward through small refinements
  until reaching executable code at the bottom. Every step preserves
  correctness; transitivity glues them. The single-space property is
  what makes the calculus elegant: there is no separate ``spec
  language'' and ``code language'' with a verification check between
  them, but one program lattice and a refinement order on it. Recent
  work by Cai et al.~\cite{cai2025} embeds Morgan's refinement laws
  into an LLM-driven code generator, using the calculus as the
  scaffolding that turns informal specifications into verifiable
  programs.

  A compiler with staged intermediate representations is doing the
  same thing less formally: higher-level dialects describe a
  computation precisely but operationally abstractly, and lowering
  makes the operational details progressively concrete. The core
  concepts of refinement calculus are --- specification, implementation,
  the refinement relation $\sqsubseteq$, refinement laws and
  monotonicity, contracts, and compositionality.

  \paragraph{Specification and implementation.}
  A specification characterises the intended behaviour of a program
  while leaving aspects of its realisation underspecified; an
  implementation resolves those choices by committing to a concrete
  execution strategy. In MLIR, \texttt{linalg.matmul}
  specifies matrix multiplication at the level of tensor semantics
  --- given two SSA tensor operands it produces a tensor containing
  their matrix product --- but leaves loop structure, reduction
  order, memory layout, and scheduling unspecified. Lowering
  through \texttt{scf}, \texttt{cf}, \texttt{memref}, and
  \texttt{vector} commits these choices and so moves the program
  down the refinement order.

  Crucially, these commitments can be made independently.
  Tiling refines the iteration space (one large loop becomes a nest
  of smaller ones) without committing to memory layout.
  Bufferization refines the data representation (tensors become
  memrefs with explicit allocation) without changing the loop
  structure. Each step is a separate refinement along a different
  axis, and compositionality (below) guarantees that composing them
  is sound. The full chain ---
  \texttt{linalg.matmul} $\to$ tiled loops $\to$ vectorized kernel
  $\to$ machine instructions --- collapses by transitivity to a
  single refinement claim: the final code satisfies the original
  specification.

  \paragraph{The refinement relation.}
  Formally, $\sqsubseteq$ is defined via weakest preconditions. For
  program $S$ and postcondition $Q$, $\mathrm{wp}(S, Q)$ is the
  weakest precondition under which $S$ is guaranteed to terminate
  in a state satisfying $Q$. Program $S_0$ is refined by $S_1$,
  written $S_0 \sqsubseteq S_1$, iff
  \begin{equation*}
    \forall Q.\ \mathrm{wp}(S_0, Q) \Rightarrow \mathrm{wp}(S_1, Q),
  \end{equation*}
  i.e.\ any guarantee provable for $S_0$ is also provable for
  $S_1$: $S_1$ may establish more, but never less. The relation is
  reflexive and transitive, so a chain $S_0 \sqsubseteq S_1
  \sqsubseteq \cdots \sqsubseteq S_n$ collapses to $S_0 \sqsubseteq
  S_n$. For the matmul example, \texttt{linalg.matmul} says ``do
  matrix multiplication''; refinement to \texttt{scf} says ``do it
  using a loop nest with explicit bounds and access patterns''. The
  refinement is sound only if the loop nest works on every input on
  which \texttt{linalg.matmul} is itself defined --- where
  \texttt{linalg.matmul} would diverge or fault (e.g.\ mismatched
  inner dimensions at run time), the refinement is allowed to do the
  same.

  Under predicate-transformer semantics, commands form a
  complete lattice ordered by $\sqsubseteq$: $\sqcap$ is demonic
  choice (greatest lower bound; both branches must be refined),
  $\sqcup$ is angelic choice (least upper bound, when it exists;
  either branch suffices), $\mathrm{abort}$ is bottom, and
  $\mathrm{magic}$ is top. Concretely, suppose $S_1$ and $S_2$ are
  both valid refinements of $S_0$. \emph{Angelic} choice $S_1
  \sqcup S_2$ would correspond to picking whichever variant is
  cheapest at run time --- but that may select a branch that
  mis-compiles a corner case the other handles, which is exactly
  what an adversarial scheduler (the ``demon'') would exploit. The
  safer reading is therefore \emph{demonic} choice $S_1 \sqcap
  S_2$, which requires correctness on all branches. MLIR's pattern
  rewriting behaves like demonic choice: any applicable pattern
  may be selected, so correctness must hold for every possible
  rewrite outcome, not only a preferred one.
  For example, two patterns that reassociate a floating-point sum
  $a + b + c$ into $(a+b)+c$ or $a+(b+c)$ are not interchangeable
  on low-precision types like E4M3FNUZ, where addition is not
  associative. Since either pattern may fire, both must be
  independently sound --- or the contract must be weakened (e.g.\
  \texttt{fastmath<reassoc>}) to permit the reordering.

  \paragraph{Monotonicity and compositionality.}
  The Refinement Calculus apparatus only works because the program-forming
  operators are monotonic in $\sqsubseteq$:
  \begin{equation*}
    S_1 \sqsubseteq T_1 \;\wedge\; S_2 \sqsubseteq T_2
    \;\Longrightarrow\; S_1; S_2 \sqsubseteq T_1; T_2,
  \end{equation*}
  and similarly for \texttt{if}, \texttt{while}, $\sqcap$, and any
  unary constructor $F$. Equivalently: if $P_1 \sqsubseteq P_2$, then
  $C[P_1] \sqsubseteq C[P_2]$ for any context $C$. This is what makes
  stepwise refinement scalable --- refining a small piece is sound
  regardless of where it sits in a larger program. Without it, a local
  improvement could be undone by its surrounding context.

  \emph{In MLIR.} The framework uses
  \texttt{unrealized\_conversion\_cast} to keep the IR well-typed
  when no direct lowering is yet available, and the bufferization
  ops \texttt{bufferization.to\_tensor} /
  \texttt{bufferization.to\_buffer} to bridge the tensor and memref
  worlds during partial conversion. Op interfaces such as
  \texttt{BufferizableOpInterface} and \texttt{TilingInterface}
  declare the local refinement obligation; pointwise containment
  of regions and ops is what would lift those local obligations
  into a global refinement claim.

  \paragraph{Contracts.}
  A \emph{contract} is a precondition $P$ on the input state and a
  postcondition $Q$ on the output state. The Hoare
  triple~\cite{Hoare1969} $\{P\}\, S\, \{Q\}$ states that if $S$
  runs from a state satisfying $P$, it terminates in a state
  satisfying $Q$. A program $S'$ \emph{implements} the contract
  when its precondition is no stronger than $P$ and its
  postcondition is no weaker than $Q$. Contracts give refinement
  an extensional definition: $S \sqsubseteq S'$ iff every contract
  satisfied by $S$ is also satisfied by $S'$.

  \emph{In MLIR.} Op verifiers run when an op is created and
  check the local part of the contract syntactically (operand and
  result type constraints, traits, structural invariants). Semantic
  obligations are pushed into op interfaces, whose method lists read
  as Hoare-style query/rewrite obligations: an op opts in by
  promising it can answer those queries consistently and produce
  the rewrite they prescribe. The dialect-conversion framework
  states the same point at the type level: ``the conversion process
  guarantees that the type contract of the IR is preserved during
  the conversion'', so ``the types of value uses will not implicitly
  change''.

  MLIR realises the structural shape of refinement calculus: a
  single-space order in which higher-level dialects sit above
  lower-level ones, contracts declared on op interfaces, and a
  type-level guarantee preserved by dialect conversion. The proof
  obligations the calculus would attach to each refinement step ---
  a witness that the lowered form satisfies the same contracts as
  the source --- live in C++ and tests rather than as
  first-class IR (\texttt{RefinementWitness}).

  \section{Abstract Interpretation}
  \label{sec:absint}
  So far we have seen term-rewriting systems
  and refinement calculus, which help ground some of the design
  principles for dialect design and lowering in MLIR. However, compilers do
  more than lower IR. They also transform and optimise it,
  and that requires analysis.

  Abstract interpretation~\cite{cousot1977,cousot1979} is a theory of
  sound static analysis --- a way to compute properties of a program
  without running it. The method is uniform: choose a property of
  interest about the concrete program (e.g.\ the integer range of
  each variable), pick a finitely-representable approximation of
  that property as an \emph{abstract domain}, and propagate values
  in the abstract domain along the program's control flow. The
  abstraction is deliberately lossy: the analysis trades precision
  for tractability so the result is computable in bounded time
  while remaining \emph{sound}.

  This is the framework that explains why dataflow analyses
  (constant propagation, interval analysis, liveness, available
  expressions, escape analysis, type inference) are correct, and how
  to trade precision for cost in a principled way. Its core
  concepts are: \textbf{concrete domain}, \textbf{abstract domain}, 
  Galois connection
  (abstraction $\alpha$, concretization $\gamma$), abstract
  transfer function, soundness, monotonicity, fixpoint, widening,
  and narrowing. 
  
  The framework's reach extends beyond conventional
  analyses: Lemerre~\cite{lemerre2023} shows that SSA translation
  can be recast as an abstract interpretation under Global
  Value Numbering. Recent work by Giacobazzi and
  Ranzato~\cite{giacobazzi2025} addresses the foundational question
  of \emph{when} an abstract interpretation can be the best
  possible on a fixed abstract domain $A$. They show that
  per-operation optimality does not in general lift to global
  optimality of the analysis composed from those per-op transfers.

  \paragraph{Concrete and abstract domains.}
  The concrete domain $D$ is the lattice of ``things the analysis is
  approximating''. Almost always $D$ is the powerset
  $\mathcal{P}(\Sigma)$, where $\Sigma$ is the set of possible program
  states (e.g.\ variable assignments). The order $\sqsubseteq$ is set
  inclusion, the join $\sqcup$ is union, the meet $\sqcap$ is
  intersection, and $\bot = \emptyset$, $\top = \Sigma$. The abstract
  domain $D^\sharp$ is a second lattice $(\sqsubseteq^\sharp,\,
  \sqcup^\sharp,\, \sqcap^\sharp,\, \bot^\sharp,\, \top^\sharp)$,
  chosen so that its elements are finitely representable and its
  operations are computable.

  Here are some examples: the interval domain (each variable's
  value is an interval $[l,u]$ with $l \in \mathbb{Z} \cup
  \{-\infty\}$ and $u \in \mathbb{Z} \cup \{+\infty\}$), the
  constant domain ($\bot$, a specific constant $c$, or $\top$), the
  sign domain ($\bot$, \texttt{neg}, \texttt{zero}, \texttt{pos},
  $\top$), and the polyhedron domain (a convex polyhedron over the
  program variables).

  Every element of $D^\sharp$ stands for a set of concrete states,
  but there are many concrete sets that no element of $D^\sharp$
  can describe exactly --- that gap is the analysis's loss of
  precision. Take the abstract sign domain ($\bot$, \texttt{neg},
  \texttt{zero}, \texttt{pos}, $\top$) as an example. The concrete
  set $\{2, 4, 6\}$ contains only positive integers and so
  abstracts to \texttt{pos}; but \texttt{pos} concretizes back to
  \emph{all} positive integers $\{1, 2, 3, \ldots\}$, so the
  ``even'' structure of the original is gone --- the sign domain
  does not encode that information in this context.
  This loss of information is not a bug but
  the analysis's deliberate trade-off: a four-element lattice
  cannot describe every subset, and the analysis pays in precision
  for what it gains in tractability.

  \paragraph{Galois connection.}
  The relationship between concrete and abstract is a pair of monotone
  maps
  \begin{align*}
    \alpha &: D \to D^\sharp
      && \text{(abstraction --- concrete set to tightest abstract description)} \\
    \gamma &: D^\sharp \to D
      && \text{(concretization --- abstract element to its concrete states)}
  \end{align*}
  required to satisfy, for every $c \in D$ and $a \in D^\sharp$,
  \begin{equation}
    \alpha(c) \sqsubseteq^\sharp a
      \quad \Longleftrightarrow \quad
    c \sqsubseteq \gamma(a).
    \label{eq:galois}
  \end{equation}

  \begin{figure}[htbp]
  \centering
  \begin{tikzpicture}[node distance=4cm]
  \node (D) {$D$};
  \node[right=of D] (Dsharp) {$D^\sharp$};
  \draw[->, thick] ([yshift=0.18cm]D.east) --
    node[above] {$\alpha$} ([yshift=0.18cm]Dsharp.west);
  \draw[->, thick] ([yshift=-0.18cm]Dsharp.west) --
    node[below] {$\gamma$} ([yshift=-0.18cm]D.east);
  \end{tikzpicture}
  \caption{The Galois connection between a concrete domain $D$
    (e.g.\ $\mathcal{P}(\Sigma)$) and an abstract domain $D^\sharp$
    (e.g.\ the interval domain). The pair $\alpha, \gamma$ is required
    to satisfy $\alpha(c) \sqsubseteq^\sharp a \iff c \sqsubseteq
    \gamma(a)$.}
  \label{fig:galois}
  \end{figure}

  This single equivalence is what is called a \emph{Galois connection}.
  It says: ``$\alpha(c)$ sits below $a$ in the abstract lattice''
  exactly when ``$c$ sits inside $\gamma(a)$ in the concrete lattice''
  --- the two viewpoints, comparing in the abstract or comparing in the
  concrete, agree. Two consequences follow and are the everyday tools:
  \begin{itemize}
    \item $\gamma \circ \alpha$ is a closure on $D$, i.e.\ $c
      \sqsubseteq \gamma(\alpha(c))$ for every $c$. Going to the
      abstract and back never gives you a smaller set; it can only
      enlarge. This is precisely the precision loss of abstraction.
    \item $\alpha \circ \gamma = \mathrm{id}_{D^\sharp}$ when the
      connection is a \emph{Galois insertion} (the usual case in
      practice): each abstract element is the unique tightest
      description of its concretization, so the round-trip is exact.
  \end{itemize}
  The Galois connection makes the phrase ``the most precise sound
  abstraction of $c$'' mathematically determinate: it is $\alpha(c)$.
  Without the Galois connection, ``best abstract description'' has no
  reference. As a concrete instance, in the interval domain
  $\alpha(\{3, 4, 5, 7\}) = [3, 7]$ --- the smallest interval
  containing every element of the set. Concretizing back gives
  $\gamma([3, 7]) = \{3, 4, 5, 6, 7\}$, which is strictly larger than
  the original set: the closure $\gamma \circ \alpha$ has invented the
  value $6$, because the interval domain cannot represent ``gaps''.
  That invented value is the precision loss, made visible. 
  
  \paragraph{Transfer functions and fixpoint.}
  Each concrete operator $F : D \to D$ is paired with an abstract
  transfer function $F^\sharp : D^\sharp \to D^\sharp$, required to
  be monotone and \emph{sound}:
  \begin{equation}
    \alpha(F(c)) \;\sqsubseteq^\sharp\; F^\sharp(\alpha(c))
      \quad \text{for every } c \in D,
    \label{eq:soundness}
  \end{equation}
  i.e.\ applying $F$ concretely and then abstracting is at most as
  informative as abstracting first and applying $F^\sharp$. A program
  with loops yields a system of monotone equations over $D^\sharp$
  whose solution is the \emph{least fixpoint}, reached by iterating
  the transfer functions from $\bot^\sharp$ until convergence. When
  the abstract domain has infinite ascending chains (e.g.\ the
  interval domain), a \emph{widening} operator forces convergence at
  the cost of precision.

  To make this concrete: consider \texttt{arith.remui \%x, \%c8}
  (unsigned remainder by 8). The concrete operator computes $x
  \bmod 8$. Its abstract transfer function on the interval domain
  ignores the input range entirely and returns $[0, 7]$ --- valid
  for any $x$. Soundness is immediate: for any concrete $x$, the
  actual result $x \bmod 8$ lies in $[0, 7]$, so the abstract
  transfer function never reports a range smaller than the truth.
  The function is imprecise (it says $[0,7]$ even when the answer is
  exactly $5$), but never wrong --- and that single per-op fact
  composes through subsequent arithmetic to prove properties of the
  whole program.

  \paragraph{Abstract interpretation in MLIR.}
  In MLIR, each abstract domain is served by its own op interface:
  an op participates in an analysis by implementing that interface's
  transfer function. For example,
  \texttt{InferIntRangeInterface::inferResultRanges} declares the
  interval-domain transfer function,
  \texttt{op->fold} serves as the transfer function for constant
  propagation (the flat constant lattice), and
  \texttt{ValueBoundsOpInterface} declares the transfer function for
  the affine-constraint domain. The framework propagates these
  transfer functions forward through the IR until a fixpoint is
  reached.

  Two interfaces illustrate the pattern concretely.
  \texttt{InferIntRangeInterface} realises the classical
  \emph{interval domain}: one $[\ell, h]$ per SSA value,
  propagated eagerly forward through the program.
  \texttt{ValueBoundsOpInterface} realises a richer
  \emph{value-bounds domain}: a system of affine (in)equalities over
  multiple index-typed values and shape dimensions, built lazily on
  demand when a transformation poses a specific query.
  The two differ in abstract domain (a single interval vs.\ a system
  of affine constraints), traversal (eager forward vs.\ lazy
  backward), and the values they reason about (any integer vs.\
  index and shape values) --- but share the same structural pattern:
  a per-op transfer function, a dialect-agnostic engine, and
  consumer passes that act on the results.

  \paragraph{Interval-domain example.}
  The fragment below uses \texttt{arith.remui} as a one-op range
  source: the result of \texttt{remui \%x, \%c8} lies provably in
  $[0, 7]$ regardless of \texttt{\%x}, because the transfer function
  for \texttt{RemUIOp} reads off the constant divisor.
  Multiplication then composes this through interval arithmetic,
  $[0, 7] \times [8, 8] = [0, 56]$. Two analysis-driven rewrites
  fire: the comparison folds because the result range collapses to
  $\{1\}$, and the second \texttt{remui} disappears because
  $[0, 56] \subseteq [0, 64)$.

\begin{lstlisting}[style=mlir]
func.func @example(%x: i32) -> (i1, i32) {
  %c8   = arith.constant 8   : i32
  %c64  = arith.constant 64  : i32
  %c100 = arith.constant 100 : i32

  %a = arith.remui %x, %c8  : i32      // a in [0, 7]
  %b = arith.muli  %a, %c8  : i32      // b in [0, 56]
  %c = arith.cmpi  slt, %b, %c100 : i32  // proves to true
  %d = arith.remui %b, %c64 : i32      // collapses to b
  return %c, %d : i1, i32
}
\end{lstlisting}

  \newpage
  After \texttt{mlir-opt --int-range-optimizations}, the body
  collapses to:

\begin{lstlisting}[style=mlir]
func.func @example(%x: i32) -> (i1, i32) {
  %c8   = arith.constant 8 : i32
  %a    = arith.remui %x, %c8 : i32
  %b    = arith.muli  %a, %c8 : i32
  %true = arith.constant true
  return %true, %b : i1, i32
}
\end{lstlisting}

  The pass propagates ranges to a fixpoint and then applies two
  rewrites: replacing any value whose range collapses to a single
  point with a constant, and eliminating $x \bmod D$ when the
  analysis proves $0 \le x < D$. Neither rewrite is reachable by
  canonicalization alone, since canonicalisers are op-local and
  cannot query cross-op dataflow results.

  \paragraph{Value-bounds example.}
  The interval domain cannot relate two values to each other ---
  it tracks each in isolation. The example below requires exactly
  that relational reasoning: a tiled loop with step-aligned bounds
  where the boundary tile size \texttt{affine.min(32, 128 - iv)} is
  provably the constant $32$. The proof uses three constraints
  simultaneously: $0 \le iv \le 96$ from \texttt{scf.for}'s
  ValueBounds implementation, $iv \equiv 0 \pmod{32}$ from the
  loop step, and $r = 128 - iv$ from \texttt{affine.apply}. The
  underlying constraint solver answers $32 \le r$ unconditionally,
  and the second argument of the \texttt{affine.min} is dropped.

\begin{lstlisting}[style=mlir]
func.func @tile_loop(%out: memref<128xf32>, %v: f32) {
  %c0   = arith.constant 0   : index
  %c32  = arith.constant 32  : index
  %c128 = arith.constant 128 : index

  scf.for %iv = %c0 to %c128 step %c32 {
    // boundary tile - provably 32 by ValueBounds
    %t = affine.min affine_map<(d0) -> (32, 128 - d0)>(%iv)
    scf.for %j = %c0 to %t step %c32 {
      %idx = arith.addi %iv, %j : index
      memref.store %v, %out[%idx] : memref<128xf32>
    }
  }
  return
}
\end{lstlisting}

  After running \texttt{mlir-opt --affine-simplify-min-max},
  the \texttt{affine.min} is replaced by the constant \texttt{\%c32}
  and the boundary special case is gone at compile time:

\begin{lstlisting}[style=mlir]
func.func @tile_loop(%out: memref<128xf32>, %v: f32) {
  %c0   = arith.constant 0   : index
  %c32  = arith.constant 32  : index
  %c128 = arith.constant 128 : index

  scf.for %iv = %c0 to %c128 step %c32 {
    scf.for %j = %c0 to %c32 step %c32 {
      %idx = arith.addi %iv, %j : index
      memref.store %v, %out[%idx] : memref<128xf32>
    }
  }
  return
}
\end{lstlisting}

  \section{Other Formal Frameworks}
  \label{sec:otherframeworks}
  \label{sec:typetheory}
  \label{sec:category}
  \label{sec:opdensemantics}

  Type theory~\cite{pierce2002,martinlof1984} studies typed
  expressions: how they are constructed, verified, and composed.  Its
  main contributions to compiler design are \emph{type classes}
  (bounded polymorphism --- a function generic over all types that
  supply a fixed interface), \emph{typing judgments} ($\Gamma \vdash e
  : T$ --- a compositional well-formedness check), and
  \emph{refinement types} ($\{x:T \mid P(x)\}$ --- a base type
  narrowed by an invariant).  Category
  theory~\cite{maclane1971,awodey2010} complements this with the
  mathematics of structure-preserving maps: objects, morphisms,
  composition, identity, and functors that translate between systems
  while preserving composition ($F(g \circ f) = F(g) \circ F(f)$).
  The framework articulates what ``structure-preserving lowering''
  means precisely and where that preservation breaks (e.g.\ the
  phase-ordering problem: transformations that commute at one
  abstraction level need not commute after lowering).

  Operational semantics~\cite{plotkin1981} defines program meaning via
  execution rules (small-step or big-step transitions); denotational
  semantics~\cite{scott1971,scott1976} maps programs to mathematical
  objects and reasons about equivalence compositionally.  The two are
  connected by adequacy and full abstraction, and together supply the
  formal ground for translation
  validation~\cite{leroy2009compcert,fehr2025}: proving that a lowered
  program denotes the same function as the source.

  \section{Reflections on Good Design}
  \label{sec:reflections}

  Each theory in this paper poses one question a compiler designer
  should be able to answer before writing code.
  Term rewriting: do your patterns reach a unique result regardless of firing order?
  Refinement calculus: does each lowering step keep the promises the previous level made?
  Abstract interpretation: what exactly is your analysis approximating, and how much precision are you giving up?
  Type theory: what contracts can types enforce so that ill-formed compositions are caught at construction time rather than at runtime?
  Category theory: what structure survives the translation between layers?

  A performant compiler that does not mis-compile answers these kinds
  of questions indirectly. But stating them in a common formal vocabulary
  makes the answers checkable rather than implicit or via extensive testing.
  It also makes the gaps visible: if you cannot state what your
  analysis is an abstraction \emph{of}, it is hard to argue that the
  analysis is sound.

  The practical payoff is not in proofs but in design clarity.
  A designer who thinks ``is my lowering monotone?'' --- meaning, if
  the input is further refined, does the output only get tighter,
  never looser --- before writing a pass will avoid a class of bugs
  that otherwise surface only under composition.
  A designer who asks ``are my canonicalisation rules
  confluent?'' will structure them so that the answer is yes by
  construction, rather than discovering non-determinism in production.

  \section{Conclusion}
  \label{sec:conclusion}
  In this paper we took a new perspective on MLIR and compiler
  design, reflecting on the correspondences between formal theories and
  implementation. The value that the formal theories surveyed here
  add is that they provide vocabulary to state what a
  design is trying to achieve, and criteria for checking whether it
  succeeds.
  As coding agents lower the cost of implementation, the
  bottleneck in compiler engineering shifts to good
  design~\cite{Lattner2026ClaudeCCompiler}.  An agent can generate a
  transformation, but it can only reason over the structure the IR
  exposes. Choosing what structure to
  expose --- what to keep abstract, what to refine, what to verify ---
  is a design problem, and it is exactly the kind of problem these
  theories help think through.

  \bibliographystyle{plain}
  \bibliography{reference_arxiv}

@inproceedings{clickir1995,
  author    = {Click, Cliff and Paleczny, Michael},
  title     = {A Simple Graph-Based Intermediate Representation},
  booktitle = {Proceedings of the ACM SIGPLAN Workshop on Intermediate Representations},
  year      = {1995},
  pages     = {35--49},
  publisher = {ACM},
  address   = {New York, NY, USA},
  doi       = {10.1145/202529.202534}
}

@online{Lattner2026ClaudeCCompiler,
  author       = {Chris Lattner},
  title        = {The Claude C Compiler: What It Reveals About the Future of Software},
  year         = {2026},
  month        = feb,
  publisher    = {Modular},
  url          = {https://www.modular.com/blog/the-claude-c-compiler-what-it-reveals-about-the-future-of-software},
  note         = {Engineering blog post},
}

@inproceedings{mlir2021,
  author    = {Lattner, Chris and Amini, Mehdi and Bondhugula, Uday and Cohen, Albert and
               Davis, Andy and Pienaar, Jacques and Riddle, River and Shpeisman, Tatiana and
               Vasilache, Nicolas and Zinenko, Oleksandr},
  title     = {MLIR: Scaling Compiler Infrastructure for Domain Specific Computation},
  booktitle = {Proceedings of the IEEE/ACM International Symposium on Code Generation and Optimization (CGO)},
  year      = {2021},
  pages     = {2--14},
  publisher = {IEEE},
  address   = {Piscataway, NJ, USA}
}

@misc{hexagonmlir2026,
  title   = {Hexagon-MLIR: An AI Compilation Stack for Qualcomm's Neural Processing Units (NPUs)},
  author  = {Absar, Mohammed Javed and Baskaran, Muthu and Sharma, Abhikrant and
             Bhandari, Abhilash and Aggarwal, Ankit and Rangasamy, Arun and
             Das, Dibyendu and Hosseini, Fateme and Slama, Franck and Brumar, Iulian and
             Verma, Jyotsna and Bindumadhavan, Krishnaprasad and Kothari, Mitesh and
             Gupta, Mohit and Kolachana, Ravishankar and Lethin, Richard and
             Narang, Samarth and Ladwa, Sanjay Motilal and Jain, Shalini and
             Dalvi, Snigdha Suresh and Rahman, Tasmia and
             Komatireddy, Venkat Rasagna Reddy and Pandya, Vivek Vasudevbhai and
             Shi, Xiyue and Zipper, Zachary},
  howpublished = {arXiv preprint arXiv:2602.19762},
  year    = {2026}
}

@misc{polyblocks2026,
  title   = {PolyBlocks: A Compiler Infrastructure for AI Chips and Programming Frameworks},
  author  = {Chen, Yuran and others},
  howpublished = {arXiv preprint arXiv:2603.06731},
  year    = {2026}
}

@misc{golin2024,
  title   = {Towards a High-Performance AI Compiler with Upstream MLIR},
  author  = {Golin, Renato and Chelini, Lorenzo and Siemieniuk, Adam and Madhu, Kavitha and
             Hasabnis, Niranjan and Pabst, Hans and Georganas, Evangelos and Heinecke, Alexander},
  howpublished = {arXiv preprint arXiv:2404.15204},
  year    = {2024}
}

@misc{lucke2024,
  title   = {The MLIR Transform Dialect: Your Compiler Is More Powerful Than You Think},
  author  = {L{\"u}cke, Martin Paul and Zinenko, Oleksandr and Moses, William S. and
             Steuwer, Michel and Cohen, Albert},
  howpublished = {arXiv preprint arXiv:2409.03864},
  year    = {2024}
}

@inproceedings{axi4mlir2024,
  title     = {AXI4MLIR: User-Driven Automatic Host Code Generation for Custom AXI-Based Accelerators},
  author    = {Agostini, Nicolas Bohm and Haris, Jude and Gibson, Perry and
               Jayaweera, Malith and Rubin, Norm and Tumeo, Antonino and
               Abell{\'a}n, Jos{\'e} L. and Cano, Jos{\'e} and Kaeli, David},
  booktitle = {Proceedings of the IEEE/ACM International Symposium on Code Generation and Optimization (CGO)},
  year      = {2024},
  publisher = {IEEE},
  address   = {Piscataway, NJ, USA}
}

@inproceedings{vasilache2022,
  title     = {Structured Operations: Modular Design of Code Generators for Tensor Compilers},
  author    = {Vasilache, Nicolas and Zinenko, Oleksandr and Bik, Aart J. C. and others},
  booktitle = {Proceedings of the International Workshop on Languages and Compilers for Parallel Computing (LCPC)},
  year      = {2022}
}

@software{iree-github,
  title  = {{IREE}: Retargetable {MLIR}-Based Machine Learning Compiler and Runtime},
  author = {{IREE Open Source Contributors}},
  url    = {https://github.com/iree-org/iree},
  year   = {2024}
}

@misc{tinyiree2022,
  title   = {TinyIREE: An ML Execution Environment for Embedded Systems from Compilation to Deployment},
  author  = {Liu, Hsin-I Cindy and Brehler, Marius and Ravishankar, Mahesh and Vasilache, Nicolas and Vanik, Ben and Laurenzo, Stella},
  howpublished = {arXiv preprint arXiv:2205.14479},
  year    = {2022}
}

@misc{aster2026,
  title        = {ASTER: MLIR C++ tool for programmable and highly-controllable assembly production on AMD GPUs.},
  author       = {{The ASTER Authors}},
  howpublished = {\url{https://github.com/iree-org/aster/tree/main}},
  year         = {2026},
  note         = {GitHub repository}
}

@misc{cudatile2026,
  title        = {CUDA Tile: A Tile-Based Programming Model and MLIR-Based Compiler Infrastructure for NVIDIA GPUs},
  author       = {{NVIDIA Corporation}},
  howpublished = {\url{https://github.com/NVIDIA/cuda-tile}},
  year         = {2024},
  note         = {GitHub repository}
}

@misc{mlir_lighthouse,
  title        = {MLIR Lighthouse Project},
  author       = {{LLVM Project}},
  year         = {2025},
  howpublished = {\url{https://github.com/llvm/lighthouse}} 
}

@misc{tvm2018,
  title   = {TVM: An Automated End-to-End Optimizing Compiler for Deep Learning},
  author  = {Chen, Tianqi and Moreau, Thierry and Jiang, Ziheng and others},
  howpublished = {arXiv preprint arXiv:1802.04799},
  year    = {2018}
}

@inproceedings{halide2013,
  title     = {Halide: A Language and Compiler for Optimizing Parallelism, Locality, and Recomputation},
  author    = {Ragan-Kelley, Jonathan and Barnes, Connelly and Adams, Andrew and others},
  booktitle = {Proceedings of the ACM SIGPLAN Conference on Programming Language Design and Implementation (PLDI)},
  year      = {2013},
  pages     = {519--530},
  publisher = {ACM},
  address   = {New York, NY, USA}
}

@misc{santos2025claudecodeconfig,
  title   = {Decoding the Configuration of AI Coding Agents: Insights from Claude Code Projects},
  author  = {Santos, Helio Victor F. and Costa, Vitor and Montandon, Joao Eduardo and Valente, Marco Tulio},
  howpublished = {arXiv preprint arXiv:2511.09268},
  year    = {2025},
  doi     = {10.48550/arXiv.2511.09268},
  url     = {https://arxiv.org/abs/2511.09268}
}

@misc{mateega2026idebench,
  title   = {IDE-Bench: Evaluating Large Language Models as IDE Agents on Real-World Software Engineering Tasks},
  author  = {Mateega, Spencer and Yang, Jeff and Costello, Tiana and Jadhav, Shaurya and Tian, Nicole and Garcinu{\~n}o, Agustin},
  howpublished = {arXiv preprint arXiv:2601.20886},
  year    = {2026},
  doi     = {10.48550/arXiv.2601.20886},
  url     = {https://arxiv.org/abs/2601.20886}
}

@inproceedings{cousot1977,
    author    = {Cousot, Patrick and Cousot, Radhia},
    title     = {Abstract Interpretation: A Unified Lattice Model for Static
                 Analysis of Programs by Construction or Approximation of Fixpoints},
    booktitle = {Proceedings of the ACM Symposium on Principles of Programming
                 Languages (POPL)},
    year      = {1977},
    pages     = {238--252},
    publisher = {ACM},
    address   = {New York, NY, USA},
    doi       = {10.1145/512950.512973}
  }

@inproceedings{cousot1979,
    author    = {Cousot, Patrick and Cousot, Radhia},
    title     = {Systematic Design of Program Analysis Frameworks},
    booktitle = {Proceedings of the ACM Symposium on Principles of Programming
                 Languages (POPL)},
    year      = {1979},
    pages     = {269--282},
    publisher = {ACM},
    address   = {New York, NY, USA},
    doi       = {10.1145/567752.567778}
  }

@article{lemerre2023,
    author    = {Lemerre, Matthieu},
    title     = {{SSA} Translation Is an Abstract Interpretation},
    journal   = {Proc. ACM Program. Lang.},
    volume    = {7},
    number    = {POPL},
    articleno = {65},
    year      = {2023},
    month     = jan,
    publisher = {ACM},
    doi       = {10.1145/3571221}
  }

@article{giacobazzi2025,
    author    = {Giacobazzi, Roberto and Ranzato, Francesco},
    title     = {The Best of Abstract Interpretations},
    journal   = {Proc. ACM Program. Lang.},
    volume    = {9},
    number    = {POPL},
    articleno = {46},
    year      = {2025},
    month     = jan,
    publisher = {ACM}
  }

@article{leroy2009compcert,
    author  = {Leroy, Xavier},
    title   = {Formal Verification of a Realistic Compiler},
    journal = {Communications of the ACM},
    volume  = {52},
    number  = {7},
    year    = {2009},
    pages   = {107--115},
    doi     = {10.1145/1538788.1538814}
  }

@techreport{plotkin1981,
    author      = {Plotkin, Gordon D.},
    title       = {A Structural Approach to Operational Semantics},
    institution = {Aarhus University},
    year        = {1981},
    number      = {DAIMI FN-19},
    note        = {Reprinted in Journal of Logic and Algebraic Programming, 60-61:17--139, 2004}
  }

@book{morgan1990,
    author    = {Morgan, Carroll},
    title     = {Programming from Specifications},
    publisher = {Prentice Hall},
    address   = {Hemel Hempstead, UK},
    year      = {1990},
    isbn      = {978-0139403576}
  }

@book{back1998,
    author    = {Back, Ralph-Johan and von Wright, Joakim},
    title     = {Refinement Calculus: A Systematic Introduction},
    publisher = {Springer},
    address   = {New York, NY, USA},
    year      = {1998},
    series    = {Graduate Texts in Computer Science},
    doi       = {10.1007/978-1-4612-1674-2}
  }

@article{fehr2025,
    author    = {Fehr, Mathieu and Fan, Yuyou and Pompougnac, Hugo and
                 Regehr, John and Grosser, Tobias},
    title     = {First-Class Verification Dialects for {MLIR}},
    journal   = {Proc. ACM Program. Lang.},
    volume    = {9},
    number    = {PLDI},
    articleno = {206},
    year      = {2025},
    month     = jun,
    publisher = {ACM},
    doi       = {10.1145/3729328}
  }

@article{cai2025,
    author    = {Cai, Yufan and Hou, Zhe and Sanan, David and Luan, Xiaokun
                 and Lin, Yun and Sun, Jun},
    title     = {Automated Program Refinement: Guide and Verify Code
                 {L}arge {L}anguage {M}odel with Refinement Calculus},
    journal   = {Proc. ACM Program. Lang.},
    volume    = {9},
    number    = {POPL},
    articleno = {69},
    year      = {2025},
    month     = jan,
    publisher = {ACM},
    doi       = {10.1145/3704879}
  }

@article{scott1976,
    author  = {Scott, Dana S.},
    title   = {Data Types as Lattices},
    journal = {SIAM Journal on Computing},
    volume  = {5},
    number  = {3},
    year    = {1976},
    pages   = {522--587},
    doi     = {10.1137/0205037}
  }

@techreport{scott1971,
    author      = {Scott, Dana S. and Strachey, Christopher},
    title       = {Toward a Mathematical Semantics for Computer Languages},
    institution = {Oxford University Computing Laboratory},
    year        = {1971},
    number      = {PRG-6},
    url         = {Please fill here}
  }

@book{pierce2002,
    author    = {Pierce, Benjamin C.},
    title     = {Types and Programming Languages},
    publisher = {MIT Press},
    address   = {Cambridge, MA, USA},
    year      = {2002},
    isbn      = {978-0262162098}
  }

@book{martinlof1984,
    author    = {Martin-L{\"o}f, Per},
    title     = {Intuitionistic Type Theory},
    publisher = {Bibliopolis},
    address   = {Naples, Italy},
    year      = {1984},
    note      = {Notes by Giovanni Sambin of a series of lectures given in Padova, June 1980},
    isbn      = {978-8870881059}
  }

@book{maclane1971,
    author    = {Mac Lane, Saunders},
    title     = {Categories for the Working Mathematician},
    publisher = {Springer},
    address   = {New York, NY, USA},
    year      = {1971},
    series    = {Graduate Texts in Mathematics},
    volume    = {5},
    doi       = {10.1007/978-1-4757-4721-8}
  }

@book{awodey2010,
    author    = {Awodey, Steve},
    title     = {Category Theory},
    publisher = {Oxford University Press},
    address   = {Oxford, UK},
    edition   = {2nd},
    year      = {2010},
    isbn      = {978-0199237180}
  }

@article{Hoare1969,
    author  = {Hoare, C. A. R.},
    title   = {An Axiomatic Basis for Computer Programming},
    journal = {Communications of the ACM},
    volume  = {12},
    number  = {10},
    year    = {1969},
    pages   = {576--580},
    doi     = {10.1145/363235.363259}
  }

@misc{merckx2026,
    title   = {E-Graphs as a Persistent Compiler Abstraction},
    author  = {Merckx, Jules and Lopoukhine, Alexandre and Coward, Samuel and
               Cheng, Jianyi and De Sutter, Bjorn and Grosser, Tobias},
    howpublished = {arXiv preprint arXiv:2602.16707},
    year    = {2026}
  }

@book{baadernipkow1998,
    author    = {Baader, Franz and Nipkow, Tobias},
    title     = {Term Rewriting and All That},
    publisher = {Cambridge University Press},
    address   = {Cambridge, UK},
    year      = {1998},
    isbn      = {978-0521779203}
  }

@article{newman1942,
    author  = {Newman, Maxwell H. A.},
    title   = {On Theories with a Combinatorial Definition of Equivalence},
    journal = {Annals of Mathematics},
    volume  = {43},
    number  = {2},
    year    = {1942},
    pages   = {223--243},
    doi     = {10.2307/1968867}
  }

@incollection{knuthbendix1970,
    author    = {Knuth, Donald E. and Bendix, Peter B.},
    title     = {Simple Word Problems in Universal Algebras},
    booktitle = {Computational Problems in Abstract Algebra},
    editor    = {Leech, John},
    publisher = {Pergamon Press},
    address   = {Oxford, UK},
    year      = {1970},
    pages     = {263--297}
  }

@article{meseguer2012,
    author  = {Meseguer, Jos{\'e}},
    title   = {Twenty Years of Rewriting Logic},
    journal = {Journal of Logic and Algebraic Programming},
    volume  = {81},
    number  = {7-8},
    year    = {2012},
    pages   = {721--781},
    doi     = {10.1016/j.jlap.2012.06.003}
  }

  \end{document}